\documentclass{desyproc}

\begin{document}
\title{Central Production of Two-Pseudoscalar Final States at COMPASS}

\author{{\slshape Alexander Austregesilo$^1$ for the COMPASS Collaboration}\\[1ex]
$^1$Technische Universit\"at M\"unchen, Physik-Department E18, 85748 Garching, Germany}

\contribID{EDSBlois/2013/36}


\acronym{EDS'13} 

\maketitle

\begin{abstract}
COMPASS is a fixed-target experiment at CERN SPS which focused on light-quark meson spectroscopy during the data-taking periods in 2008 and 2009. The central exclusive production of glueball candidates is studied with a 190\,GeV/$c$ proton beam impinging on a liquid hydrogen target. We select centrally produced systems with two pseudo-scalar mesons in the final state. The decay of this system is decomposed in terms of partial waves, with particular attention paid to the inherent mathematical ambiguities of the amplitude analysis. We show that simple parametrisation are able to describe the mass dependence of the fit results with sensible Breit-Wigner parameters.
\end{abstract}

\section{Introduction and Kinematic Selection}

\begin{figure}[ht]
    \begin{minipage}{.36\textwidth}
      \begin{center}
        \vspace{.5cm}
        \includegraphics[width=\textwidth]{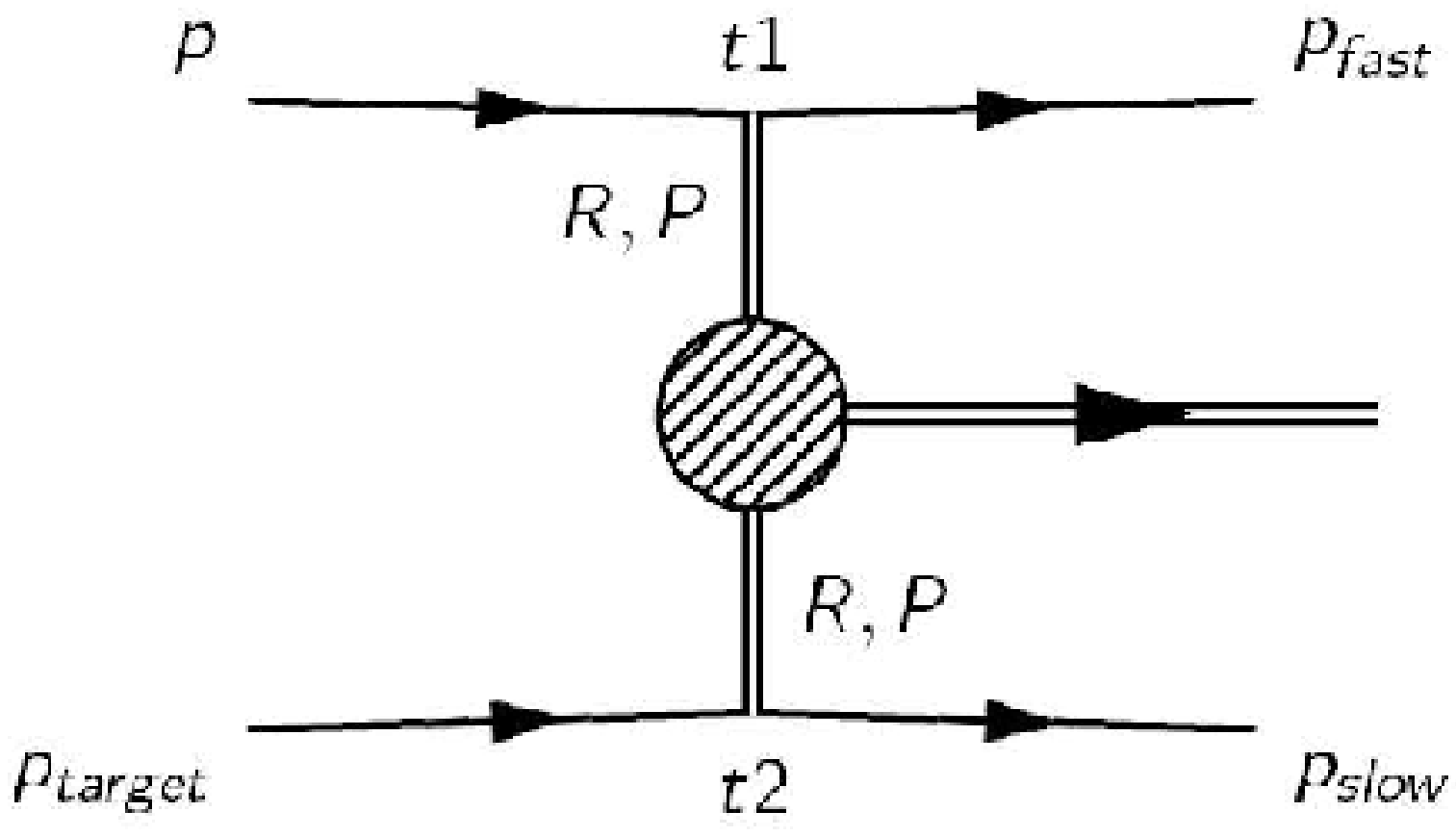}
        \caption{\em  Central production.}
        \label{fig:cp}
      \end{center}
    \end{minipage}
    \begin{minipage}{.31\textwidth}
      \begin{center}
        \includegraphics[width=\textwidth]{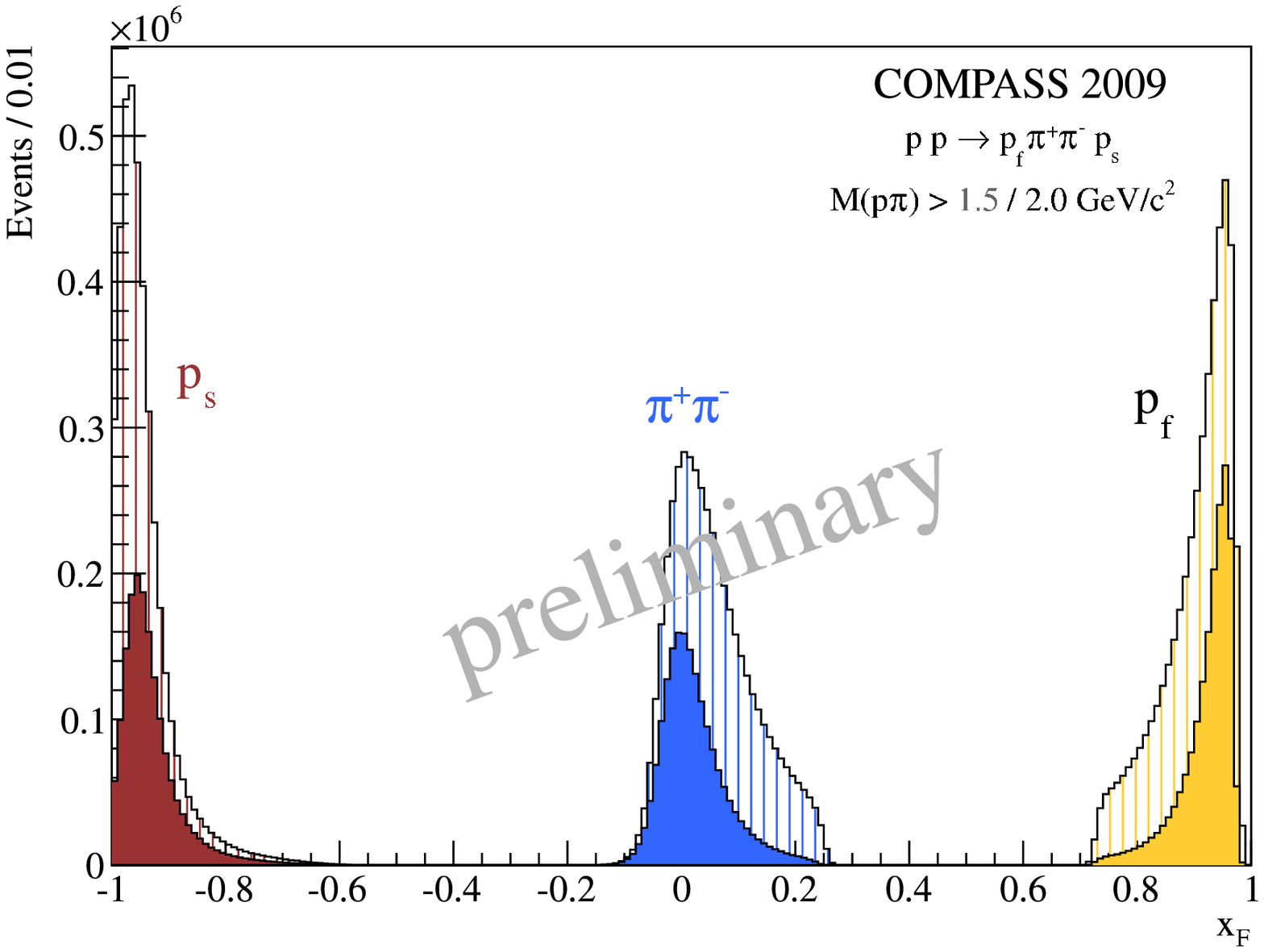}
        \caption[{\em Feynman $x_F$}]{\em  Feynman $x_F$.}
        \label{fig:xf}
      \end{center}
    \end{minipage}
    \begin{minipage}{.31\textwidth}
      \begin{center}
        \includegraphics[width=\textwidth]{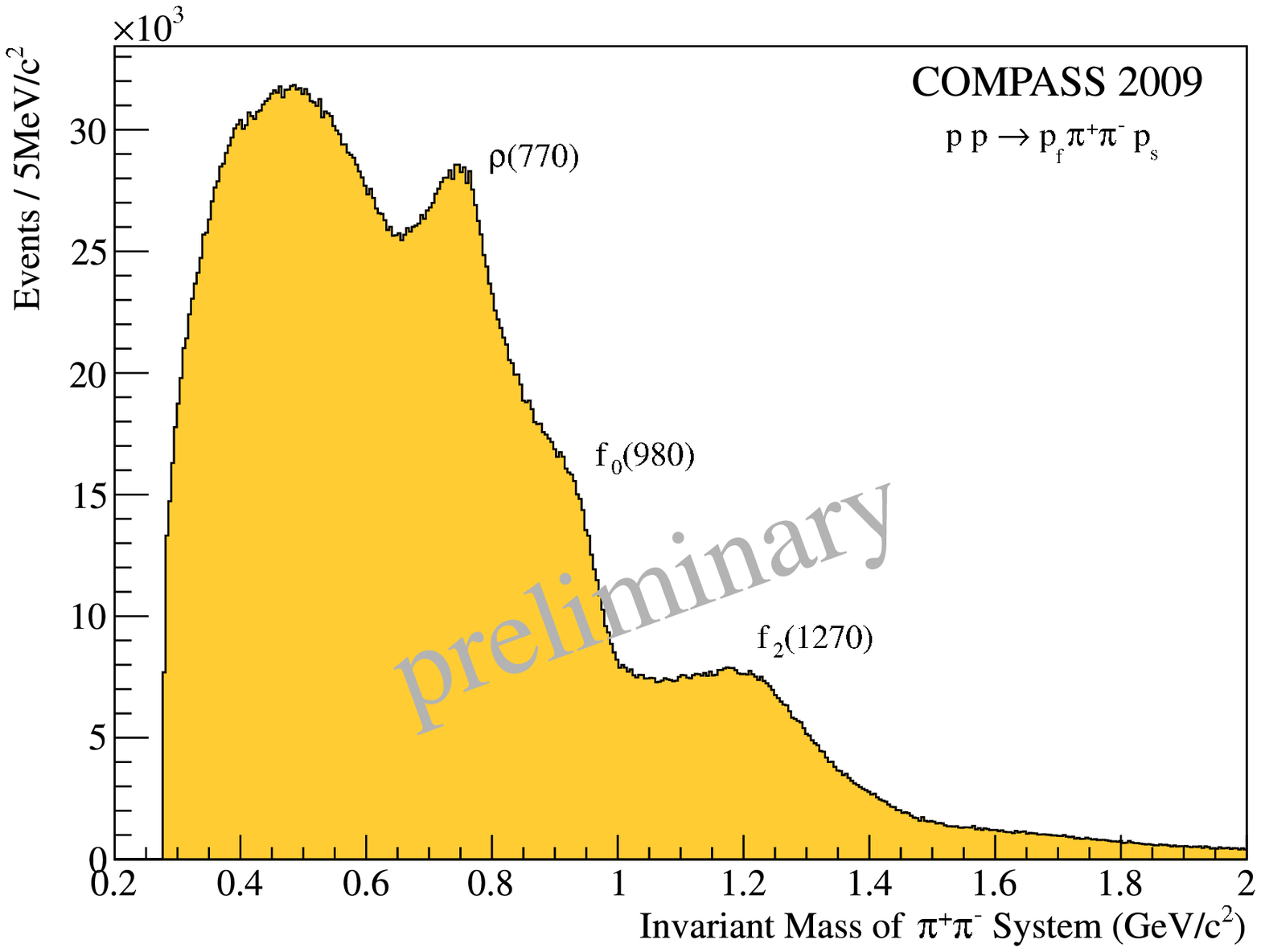}
        \caption[{\em $\pi^+\pi^-$ Invariant Mass}]{\em $\pi^+\pi^-$ system.}
        \label{fig:M}
      \end{center}
    \end{minipage}
 \end{figure}

One of the goals of the COMPASS experiment~\cite{com07} at CERN is to study the existence and signatures of glueballs in the light-quark sector via central exclusive production. It is realised in the fixed-target experiment by the scattering of a proton beam on a proton target, where a system of particles is produced at central rapidities (cf. Figure~\ref{fig:cp}).

A $190\,\textrm{GeV}/c$ proton beam impinging on a liquid hydrogen target was used to collect the discussed data set. The trigger on the recoil proton created a large rapidity gap between the slow proton $p_s$ and the other final-state particles measured in the forward spectrometer. Additional kinematic cuts where used in order to separate the central two-pseudoscalar system from the fast proton $p_f$. For the di-pion system, for example, a cut on the invariant mass combinations $M(p\pi) > 1.5\,\mathrm{GeV}/c^2$ was introduced. The effect of this selection on the Feynman $x_F$ distributions is shown in Figure~\ref{fig:xf}. The $\pi^+\pi^-$ system lies well within $\left \vert x_F \right \vert \le 0.25$ and can therefore be considered as centrally produced.

Figure~\ref{fig:M} shows the invariant mass of the central $\pi^+\pi^-$ system, where the $\rho$(770), the $f_2$(1270), and the sharp drop in intensity in the vicinity of the $f_0$(980) resonance can be observed as dominant features. Since the $\rho$(770) cannot be produced via double-Pomeron exchange (DPE), contributions from other production mechanisms are evidently non-negligible at $\sqrt s = 19\,\mathrm{GeV}/c^2$. A variety of selection criteria have been used in the past, and several of them have been studied with our data set. Most of them showed a large overlap, but none was able to successfully single out a clean DPE sample. However, the results discussed below exhibit little dependence on this choice.

\section{Partial-Wave Analysis in Mass Bins}

The partial-wave analysis has been performed assuming that the central two-pseudoscalar system is produced by the collision of two particles which are emitted by the scattered protons and which form the $z$-axis in the centre-of-mass frame of the $\pi^+\pi^-$ system. The $y$-axis of the right-handed coordinate system is defined by the cross product of the momentum vectors of the two exchange particles in the $pp$ centre-of-mass system. Apart from the invariant mass, the polar and azimuthal angles $\cos\theta$ and  $\phi$ of the negative particle, measured in the two-pseudoscalar centre-of-mass frame relative to the axes defined above, characterise the decay process.

 With complex transition amplitudes, the intensity in narrow mass bins can be expanded in terms of spherical harmonics. An extended maximum-likelihood fit in $10\,\mathrm{MeV}/c^{2}$ mass bins is used to find the amplitudes, such that the acceptance corrected model matches the measured data best. The transition amplitudes are constant over the narrow mass bins, which means no mass dependence is assumed at this stage.

As shown in \cite{sad91}, this decomposition of the intensity for two-pseudoscalar final states generally inhibits intrinsic ambiguities, which can be addressed with the method of Barrelet-zeros~\cite{bar72, chu97}. The mathematically equivalent solutions can be uniquely identified, however, differentiation among them requires additional input, e.g. the behaviour at threshold or the expected physical content. However, the choice is not evident for the $\pi^+\pi^-$ system. Four solutions can be clearly ruled out, since most of the intensity is formed by one single wave. The other four solutions remain subject to further studies, where different final states and the mass-dependent parametrisation may bring extra input.

  If we apply the same analysis technique to the centrally produced $K^+K^-$ system, the choice of the single physical solution becomes clearer. Only for one solution, the expected dominance of the $S$-wave at threshold is observed. In addition, this solution shows almost no intensity in the $P$-wave above the narrow $\phi$(1020), a fact that supports the assumption of double-Pomeron exchange as the dominant production process. For this reason, we limit the analysis to spin-0 and spin-2 contributions in the interesting mass range above $1.05\,\mathrm{GeV}/c^2$.

\section{Parametrisation of Mass-Dependence of $K^+K^-$} 

\begin{figure}[]
   \begin{center}
     \includegraphics[width=.85\textwidth]{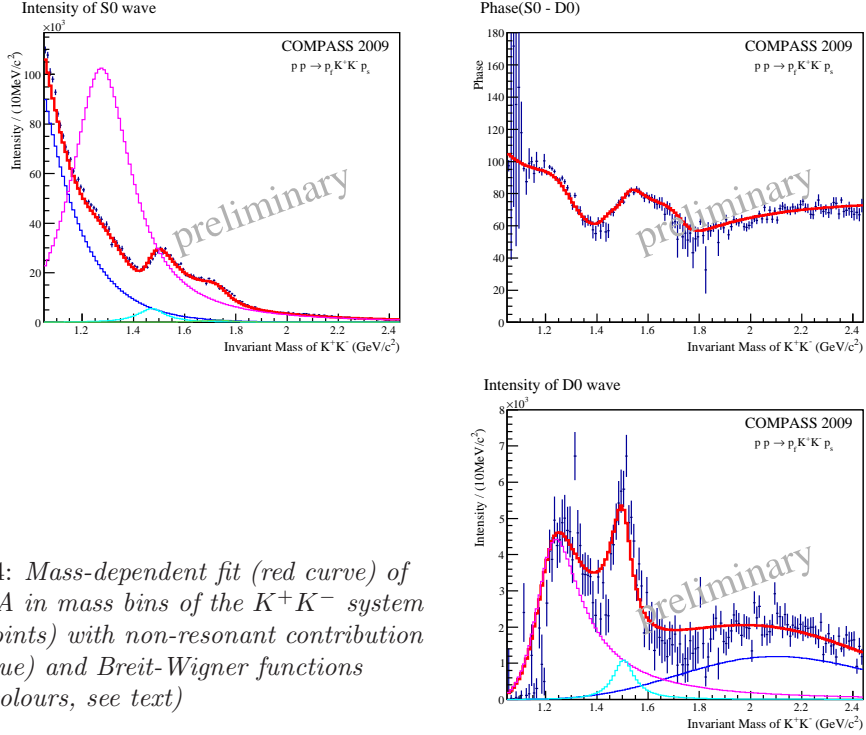}
   \end{center}
   \vspace{-3.3cm}
   \caption[{\em Mass-dependent parametrisation of intensities and phase}]{\em Mass-dependent fit (red curve) of \\ the PWA in mass bins of the $K^+K^-$ system \\ (data points) with non-resonant contribution \\ (dark blue) and Breit-Wigner functions \\ (other colours, see text)}
   \label{fig:intSD}
 \end{figure}

As a second step, the mass dependence of the partial-wave intensities and their interference is parametrised in terms of a physical model. The model parameters are determined by a $\chi^2$-fit to both the real and imaginary parts of the spin-density matrix elements. In this preliminary analysis, we focused only on the two most prominent contributions: the real (anchor) wave $S_0^-$ and the complex-valued $D_0^-$. The resonant contributions are modelled with dynamic-width relativistic Breit-Wigner functions.
Besides the resonance parameters $m_0$ and $\Gamma_0$, we allow for a free complex amplitude for every Breit-Wigner function. A non-resonant contribution has to be introduced to account for the threshold enhancement in the $S$-wave as well as for the contributions from other production processes which translate through their angular characteristics into intensity in the $D$-wave at masses above $1.8\,\mathrm{GeV}/c^2$. It is modelled by a two-body phase space with the  correct asymptotic behaviour for the angular-momentum barrier and an exponential damping for high masses. The final result can be seen in Figure~\ref{fig:intSD}.

 The two sharp peaks in the $D$-wave intensity were fitted with two well-known resonances. Breit-Wigner functions are used to parametrise the $f_2$(1270) and the $f_2'$(1525) mesons. An additional $f_2$(2150) as suggested by \cite{bar99} was not needed in order to describe the data, the intensity can be attributed to the non-resonant model component. At least three different Breit-Wigner functions 
 were necessary to describe the $S$-wave. In addition to the well-established $f_0$(1500) and $f_0$(1710) resonances, a broad $f_0$(1370) had to be included to account for both the intensity as well as the phase with respect to the $D$-wave in this mass region. A strong interference with the non-resonant contribution even required a dominant contribution from this term~(cf.~Figure~\ref{fig:intSD}). However, the large correlation between the parameters of the $f_2$(1270) in the $S$-wave and the $f_0$(1370) in the $D$-wave may lead to systematic errors in the Breit-Wigner parameters which are still under investigation. In addition, the $f_0$(1370) strength is very sensitive to the background parametrisation. For that reason, we do not quote the mass and width parameters of the resonant contributions here.

\section{Conclusion and Outlook}

We are able to select centrally produced two-pseudoscalar final states and describe the main features of the data in terms of partial waves. A more detailed description of the analysis methods can be found in~\cite{aus13}.

The amount of data as well as the sensitivity of the analysis largely exceed earlier studies. The phase relations emerge with unprecedented precision and provide important information ignored by previous analyses~(e.g.~\cite{bar99}). The data show that COMPASS may be able to contribute to the controversial discussion about the existence of resonances in the scalar sector~\cite{och13}. In order to interpret the composition of the super-numerous scalar resonances, a combined analysis of all available final states will be essential. Especially the combination with the corresponding neutral final states $\pi^0\pi^0$, $\eta\eta$, and $K^0_SK^0_S$ can help to resolve remaining ambiguities.

\section{Acknowledgements}

The author acknowledges financial support by the German Bundesministerium f\"ur Bildung und Forschung (BMBF), by the Maier-Leibnitz-Laboratorium der LMU und TU M\"unchen, and by the DFG cluster of excellence 'Origin and Structure of the Universe'.


\begin{footnotesize}
  
\end{footnotesize}
\end{document}